\documentclass[twocolumn, secnumarabic, amssymb, noshowpacs,nobibnotes, aps, showpacs,prb]{revtex4-1}
\usepackage{graphics}
\usepackage[dvips]{graphicx}
\usepackage{dcolumn}
\usepackage{amsmath}
\usepackage{lipsum}

\begin{document}
\title{Theory of finite periodic systems: The eigenfunctions symmetries}
\author{Pedro Pereyra}
\address{F\'{i}sica Te\'{o}rica y Materia Condensada, UAM-Azcapotzalco, Av. S. Pablo 180, C.P. 02200, M\'{e}xico D. F., M\'{e}xico }
\date{\today}

\begin{abstract}
Using the analytical expressions for the genuine eigenfunctions  $\varphi_{\mu\nu}(z)$ and eigenvalues $E_{\mu,\nu}$, of open, bounded and quasi-bounded finite periodic systems, we derive the eigenfunctions space-inversion symmetry relations. The superlattice eigenfunctions symmetries, closely related with the symmetries and zeros of the Chebyshev polynomials of the second kind $U_n$, are fully written in terms of the number of unit cells $n$, the subband index $\mu$ and the intra-subband index $\nu$.
\end{abstract}

\maketitle

\section{Introduction}
The eigenfunctions and eigenvalues derived within the finite periodic system theory, for open (scattering), bounded and quasi-bounded superlattices,\cite{PereyraCM2000,Pereyra2002,Pereyra2005} are the genuine quantum solutions for the actual Maxwell and Shr\"odinger equations for periodic systems. The explicit expressions obtained for the energy eigenvalues and eigenfunctions have been successfully applied to study a number of physical systems where particles and waves pass through, or become confined giving rise to important physical transitions and phenomena.\cite{Yeh1998,PereyraSimanjuntak2007,Hsueh2007,Pacher2007,Li2008, Pereyra2009,Pfeffer2009,
Kuo2009,PereyraAvila2012} Our purpose here is to discuss and determine the  parity symmetries for the resonant functions and for the eigenfunctions of open and bounded systems, respectively.

The superlattices (SLs),  introduced in the early days of the quantum theory to describe `peculiar' metallic crystals,\cite{Seemann1930} has had a large evolution,\cite{Peierls1936,Wilson1938,Wang1945,Sato1961} and many properties, among them the electronic states in long-period SL alloys and the propagation of electromagnetic waves through periodic media, have been studied.\cite{Jones1969,Abeles,BornWolf,Shik1973,Kazarino1973,Jones1973,Segmuller1973,Ludeke1974,
Shik1975Review,Dohler1975,Dohler1984,Fasolino1984,Kubota1984} The development of  techniques to grow semiconductor heterolayers, first suggested  by Kroemer\cite{Kroemer1957} and intensively  pursued in the race to achieve better optical devices,\cite{Kroemer2,Alferov1967,Ruprecht1967,Alferov1970,PanishHayashi1970} reached in the 70's the ability to produce semiconductor SLs.\cite{Keldysh1962,EsakiTsu1970}  The interest in semiconductor SLs grew immensely after the fundamental contributions on optoelectronics and resonant tunneling.\cite{Kazarinov1971,Alferov1971,Esaki1972,Tsu1973,Chang1974,Dingle1974,Mukherji1975}

The impact of the development of semiconductor SLs  has been enormous. It opened the possibility of building an unlimited variety of  artificial periodic systems; at the same time it has repositioned a fundamental problem in the quantum theory, the problem of solving the Schr\"odinger equation of periodic systems, {\it taking into account the finiteness nature as a fundamental requisite}. With SLs, the finiteness characteristic becomes more evident than ever. A SL, or a multiple-quantum well structure, is conceptually a system closer to simple structures like the quantum well, the double-quantum barrier, etc. With SLs, the theoretical descriptions  in terms of concepts valid for infinite periodic systems, became less appropriate at the same time that an alternative theory, the theory of finite periodic systems (TFPS), emerged.\cite{Claro1982,Ricco,Vezzetti, RPA, Lee, Kolatas, Griffiths, Sprung, Rozman, PereyraPRL,Yeh1998,Pereyra1998,PereyraCM2000,Pereyra2002,Pereyra2005} The TFPS is a theory that rests on the basis of a single-cell quantum solution, while the standard  theory  rest on Bloch's theorem.\cite{Bloch,Brillouin1953} This theorem and the ensuing Bloch function, rigorously valid for {\it infinite} systems, were crucial for the development of solid state physics and the theoretical description of macroscopic crystalline systems, which, generally, contain a large number of unit cells.  Nevertheless, the appearance of SLs is leading, steadily, to introduce a different but suitable approach, appropriate not only for these systems but also for simpler crystalline structures.

Given the explicit representations of eigenfunctions of finite-periodic systems and the increasing use of superlattices in the active zone of laser devices,  it is interesting and useful to study, analytically, and determine the space-inversion symmetries of these functions. The parity relations derived here are particularly appropriate for layered structures, from double barrier to superlattices. After a brief outline of the TFPS we will derive, in section \ref{SecTFPS}, the symmetry relations of the Chebysheb polynomials and those of the resonant states in open superlattices. In section 4, we deal with SLs bounded by hard walls and the symmetries of their eigenfunctions. Finally, in section 5, we derive the space-inversion symmetry relations of the eigenfunction of SLs  bounded by lateral barriers with finite height, and we will end up with some conclusions.

\section{fundamentals of the theory of finite periodic systems}\label{SecTFPS}

At variance with the  attempts to adapt the standard  theory to finite lattice descriptions, using forces  and boundary conditions,\cite{Brillouin1953} a different approach,  mathematically simpler and conceptually neat, with roots in the theoretical approaches to optical waves in periodic structures and the electronic transport through disordered conductors,\cite{Abeles,BornWolf,Weinstein,Jakobson,Yarib, Erdos,Mello,MPK,Pereyra1995}  has grown and gradually developed into the actual theory of finite periodic systems,  with applications in a great diversity of topics.\cite{Pereyra2000,Griffiths2001,Assaoui2002,Kunold2003,Pacher2003,Simanjuntak2003,
Koehler2005,Diago,Pacher2007,Peisakovic2008,Pfeffer2009,Pereyra2011}  The original aim of calculating transport properties through a simple one-dimensional periodic potential, has evolved into the theoretical approach to describe the transport through periodic structures with arbitrary number $n$ of unit cells, arbitrary number $N$ of physical channels or propagating modes and arbitrary potential profiles, and for the calculation of energy eigenvalues and the corresponding eigenfunctions.  The TFPS, is based on the transfer matrix method, which properties allow to write the n-cells transfer matrix $M_n$ as $M^n$, with $M$ the single-cell transfer matrix. This relation has been then transformed into the non-commutative recurrence relation\cite{Pereyra1998,PereyraPRL}
\begin{equation}\label{RecRelation}
p_n-(\beta^{-1} \alpha \beta+ \alpha^*)p_{n-1}+p_{n-2}=0.
\end{equation}
where $\alpha$ and $\beta$ are the (1,1) and (1,2) blocks of the 2$N\times 2N$ transfer matrix $M$. This problem was solved and matrix polynomials $p_n$ of dimension $N$$\times$$N$ were obtained.\cite{Pereyra1998,PereyraPRL} In terms of the matrix polynomials $p_n$, the transmission amplitudes are given by
\begin{equation}\label{TransmCoeffGen}
t_n^T=(p_n-\beta^{-1} \alpha \beta p_{n-1})^{-1},
\end{equation}
In the widely used one-dimensional, one-propagating mode limit, equation (\ref{RecRelation}) is the recurrence relation of the well known Chebyshev polynomial of the second kind, $U_n$, evaluated at the real part of $\alpha=\alpha_R+i\alpha_I$. In this case, the transmission amplitude becomes
\begin{equation}\label{TransmCoeff}
t_n=(U_n-\alpha U_{n-1})^{-1}.
\end{equation}
This result known in the electromagnetic theory, for stratified media,\cite{BornWolf, Abeles,Jakobson,Yarib} was  rediscovered in the 1D approaches for electronic transport through layered structures.\cite{Claro1982,Ricco,Vezzetti, RPA, Lee, Kolatas, Griffiths, Sprung, Rozman, PereyraPRL} Besides the resonant levels, wave functions and transmission coefficients of open SLs,  {\it bona fide} eigenvalues and eigenfunctions, for bounded and quasi-bounded 1D periodic systems, were also  obtained.\cite{PereyraCM2000,Pereyra2002,Kunold2003,Pereyra2005,Pacher2007}

When SLs are used in the active zone of light emitting devices, an important quantity to calculate is the optical response based on the golden rule
\begin{eqnarray}
|\langle\psi_{\rm f} |H_{\rm int}|\psi_{\rm i}\rangle |^2/[E_{\rm f}-E_{\rm i}+\hbar \omega)^2+\Gamma_{\rm i}^2],
\end{eqnarray}
where $H_{\rm int}$ represents the light-matter interaction, $|\psi_{i}\rangle$ and $|\psi_{f}\rangle $ the initial and final states,  $E_{i}$ and $E_{f}$ the corresponding energies and $\Gamma_{i}$ the decay rate. To calculate explicitly  the optical response it is necessary to know the SL eigenfunctions and the correspondent eigenvalues. How is then that not knowing explicitly the eiegenfunctions and eigenvalues, the optical responses have been so far calculated. The matrix elements are replaced by the so-called oscillator strengths determined, generally, through  indirect and rather cumbersome procedures.

In the theory of finite periodic systems, each subband is characterized by a set of eigenfunctions $\phi_{\mu \nu}(z)$ and eigenvalues $E_{\mu \nu}$.\cite{PereyraCM2000,Pereyra2002,Pereyra2005} In the numerical calculation of transition probabilities and the evaluation of the large number of integrals like
\begin{eqnarray} \int dz
\varphi^{v}_{\mu^{'},\nu^{'}}(z)\frac{\partial}{\partial z}\varphi^{c}_{\mu,\nu}(z),
\end{eqnarray}
the eigenfunction symmetries play a fundamental role. The superindices $v$ and $c$ stand for valence and conduction bands. Our purpose here is to consider the analytical expressions obtained in Ref. [\onlinecite{Pereyra2005}] for the resonant functions and the eigenfunctions of open, and bounded and quasi-bounded superlattices, and to determine their space-inversion symmetries that will support the optical-transition selection rules. We will show that each eigenfunction  $\phi_{\mu \nu}(z)$  has a well defined symmetry, determined by the number of unit cells $n$,  the subband index  $\mu$ and the intra-subband index $\nu$.

\section{Symmetries of Chebyshev polynomials and the resonant states in open periodic systems}\label{SymChPol}

\begin{figure}\label{Fig1}
\begin{center}
\includegraphics [width=230pt]{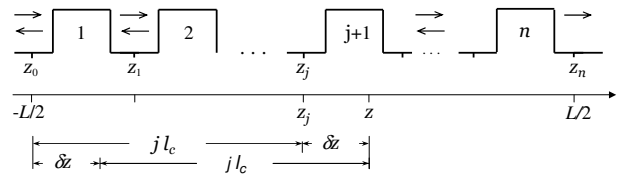}
\caption{An open SL with $n$ unit cells. The coordinate $z$ at any point in the $j+1$ cell, (where $j=$1, 2, ...,n),  can be written as $z=j l_c+\delta z$.}
\end{center}
\end{figure}

\begin{figure}[b]
\begin{center}
\includegraphics [width=200pt]{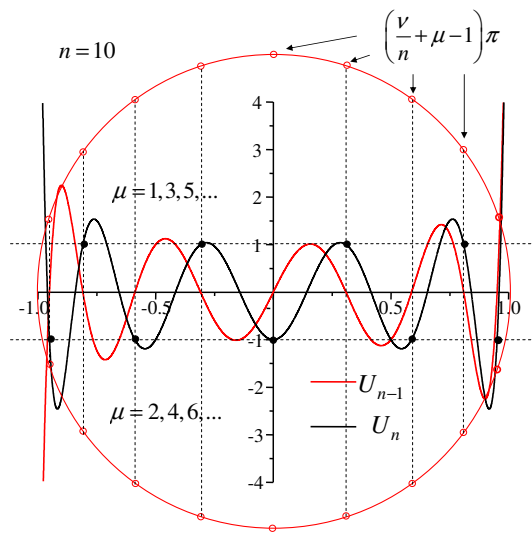}
\caption{The zeros of the Chebyshev polynomial $U_{n-1}$ coincide with $\cos (\nu\pi/n +(\mu-1)\pi )$, where $\nu,\mu$$=1,2,3,..$, are the intrasubband and subband indices, respectively. At these points $U_n$=$\pm$ 1 (see black points).}\label{Fig2}
\end{center}
\end{figure}
We will assume that a $n$-cell system is connected to ideal leads. Even though the results that will be obtained here are valid in general, i.e. for any specific shape of the single cell potential profile, we will consider in this section and the coming ones, SLs with sectionally-constant potential, as shown in figure 1,  known also as the finite Kronig-Penney model.

A well know feature of the transmission coefficients, of finite periodic structures, is the resonant behavior. The transmission coefficient $T_n$=$|t_n|^2$ is 1 whenever the incoming energy coincides with the resonant energies. It was shown in the theory of finite periodic systems that, independently of the specific potential shape, the transmission coefficient $T_{n}$ can be written as
\begin{equation}
\left| t_{n}\right| ^{2}=\frac{\left| t\right| ^{2}}{\left| t\right|
^{2}+U_{n-1}^{2}(1-\left| t\right| ^{2})}, \label{e.3}
\end{equation}
where  $\left| t\right| ^{2}=1/\left| \alpha \right| ^{2}$ is the
single-cell transmission coefficient.   It is clear from this
equation that the transmission resonances occur precisely when the
polynomial $U_{n-1}$ becomes zero. This result\cite{Eq} could also be inferred from equation (\ref{TransmCoeff}) and the Chebyshev polynomials properties shown in figure  \ref{Fig2}, where the Chebyshev polynomials $U_n$ and $U_{n-1}$ are plotted for $n=10$. In this figure we see, on one side, that the points $x_{\nu}$ where $U_{n-1}$ is zero, coincide with the cosine of $(\nu +(\mu-1) n )\pi /n$ for $\nu=$1,2,3,...$n-$1 and $\mu=$1,2,3,..., where, on the other side, the Chebyshev polinomial $U_n$ is either +1 or $-$1. Therefore, the points $x_{\nu}$ define the conditions of complete transmission. Consequently, the $\nu$-th resonant energy in the $\mu $-th band  is the solution of
\begin{equation}\label{e.7}
(\alpha _{R})_{\mu,\nu }=\cos \frac{\nu +(\mu-1) n }{n}\pi.
\end{equation}

Solving this equation we obtain  the  whole set of resonant energies $\{E_{\mu ,\nu }\}$, which are some times denoted as  $\{E_{\mu ,\nu }^*\}$. We will not use this notation to avoid confusion with the complex conjugation.  Given these energies the corresponding resonant states  are straightforwardly obtained. An older report of this kind of functions appear in Ref. [\onlinecite{Yeh1998}]; we shall however consider the resonant functions reported in Ref. [\onlinecite{Pereyra2005}]. The state vector at any point $z$ inside the $j+1$ cell in figure 1, using the transfer matrix definition applied to this system,
becomes
\begin{equation}
\Phi (z) = M(z,z_o) \Phi (z_o)  =  M_{p}(z,z_j)M_{j}(z_j,z_o)  \left( \begin{array}{c}
a_o \cr
b_o
\end{array}
\right).  \label{e.8}
\end{equation}
Here
\begin{equation}
M_{j}(z,z_j)=\left(
\begin{array}{cc}
\alpha _{j} & \beta _{j} \\
\beta _{j}^{\ast } & \alpha _{j}^{\ast }
\end{array}
\right),
\end{equation}
is the transfer matrix that connects the state vector at the left side of the SL with the state vector at $z_j=z_o+j l_c$, thus $\alpha _{j}=U_{j}-\alpha ^{\ast }U_{j-1}$, $\beta _{j}=\beta
U_{j-1}$.  $M_{p}$ is the transfer matrix that connects the state vectors at $z_j$ and at $z=z_j+\delta z$. If we write this matrix as
\begin{equation}
M_{p}(z,z_j)=\left(
\begin{array}{cc}
\alpha _{p} & \beta _{p} \\
\gamma _{p} & \delta _{p}
\end{array}
\right)
\end{equation}
and we assume that the incidence is only from the left hand side, we have
\begin{equation}
b_o=-\frac{\beta _{n}^{\ast }}{\alpha _{n}^{\ast
}}a_o=r_na_o,
\end{equation}
where $\alpha _{n}$ and $\beta _{n}$ are the  matrix elements of the transfer matrix $M_n$, $a_o$ is a normalization constant and
$r_n$ the total reflection amplitude; on the other side, the wave function can be written as
\begin{eqnarray}\label{OpenSLWF}
\Psi (z,E) &=& \Bigl[(\alpha _{p}+\gamma _{p}) \Bigl(\alpha _{j}-\beta _{j}\frac{\beta _{n}^{\ast }}{\alpha _{n}^{\ast }}\Bigr) \nonumber \\
&& + (\beta _{p}+\delta _{p}) \Bigl(\beta _{j}^{\ast }-\alpha _{j}^{\ast }\frac{\beta _{n}^{\ast
}}{\alpha _{n}^{\ast }}\Bigr)\Bigr]a_o
\end{eqnarray}
with the resonant wave functions given by
\begin{equation}
\Psi _{\mu ,\nu }(z)=\Psi (z,E_{\mu ,\nu }). \label{e.10}
\end{equation}
\begin{figure}
\begin{center}
\includegraphics [width=210pt]{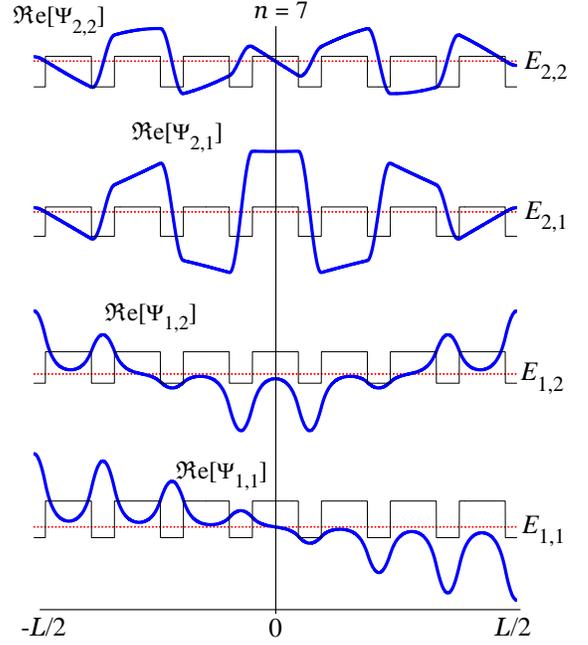}
\caption{Real parts of the resonant wave functions $\Psi _{1 ,1 }$,  $\Psi _{1 ,2 }$, $\Psi _{2 ,1 }$ and $\Psi _{2 ,2 }$, for an open SL with  $n$=7. In this case the parity $P\bigl[\mathfrak{R}{\rm e}[\Psi _{\mu ,\nu }]\bigl]$ is given by  $(-1)^{\nu+\mu+1}$. See Eq. (\ref{SimmPolinUn}). For this graph and for the following ones, we consider that the barrier height is $V_b=0.4165$eV, the well width $a$=3nm and the barrier width $b$=6nm. } \label{f1}
\end{center}
\end{figure}
In order to determine the space-inversion symmetries of these functions, we will consider pairs of points, symmetric  with respect to the middle point of the SL, that will coincide with the z-axis origin. We  will   explicitly evaluate the eigenfunctions at those points  and derive the symmetry relations. Let the points  $z_n=L/2$ and at $z_o=-L/2$. At these points, $\alpha_p=\delta_p=1$ and $\beta_p=\gamma_p=0$. At $z_o$,  $\alpha_j=1$ and $\beta_j=0$, but at  $z_n$,  $\alpha_j=\alpha_n$ and $\beta_j=\beta_n$. Taking into account that $\alpha_n\alpha_n^*-\beta_n\beta_n^*=1$, we have
\begin{eqnarray}
\Psi_{\mu,\nu}(-L/2)=(1-\frac{ \beta _{n}^{\ast }}{\alpha _{n}^{\ast }})
a_o= (1+r_{n})
a_o
\end{eqnarray}
\begin{eqnarray}
\Psi_{\mu,\nu}(L/2)=\frac{ 1}{\alpha _{n}^{\ast }} a_o = t_{n} a_o
\end{eqnarray}
Since
\begin{eqnarray}
t_n\Bigl |_{E=E_{\mu,\nu}}=\frac{1}{U_n}\Bigl |_{E=E_{\mu,\nu}}\hspace{0.2in} \text{and}\hspace{0.2in} r_n\Bigl |_{E=E_{\mu,\nu}}=0,
\end{eqnarray}
we end up with the relation
\begin{eqnarray}\label{OpenWFSymms1}
\Psi_{\mu,\nu}(L/2)=\frac{1}{U_n}\Psi_{\mu,\nu}(-L/2),
\end{eqnarray}
which shows that the parity of the resonant eigenfunctions depends on the Chebychev polynomial $U_n$ evaluated at the resonant energies. As before equation  (\ref{e.7}), a simple analysis, supported by figure \ref{Fig2}, shows that
 \begin{eqnarray}\label{SimmPolinUn}
\frac{1}{U_n}\Bigl |_{E_{\mu,\nu}}\!=\!\Biggl\{\begin{array}{cc} (-1)^{\nu} & \text{for}\hspace{0.1in} n \hspace{0.1in}\text{even}\cr  & \cr (-1)^{\nu+\mu+1} & \text{for}\hspace{0.1in} n \hspace{0.1in}\text{odd} . \end{array}\Biggr.
\end{eqnarray}
These are important symmetries that will be used below.
From here on, and to simplify the notation, it will be understood that whenever  $\Psi_{\mu,\nu}$ appears, the other quantities are also evaluated at $E=E_{\mu,\nu}$.

\begin{figure}
\begin{center}
\includegraphics [width=210pt]{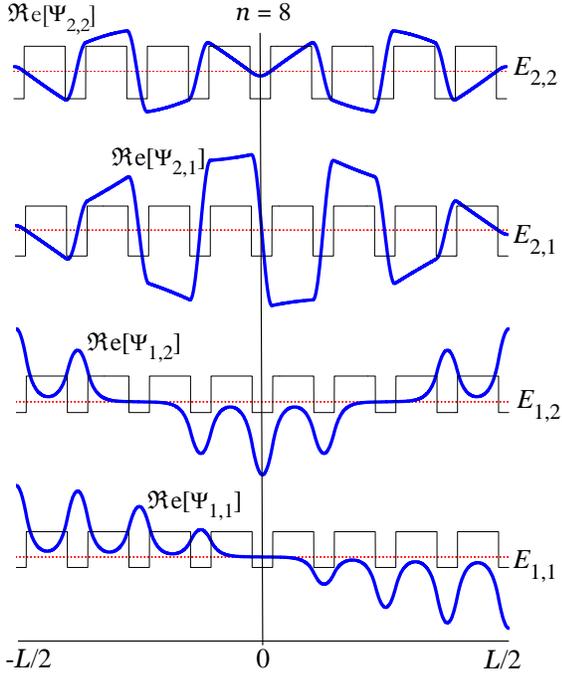}
\caption{Real parts of the resonant wave functions $\Psi _{1 ,1 }$,  $\Psi _{1 ,2 }$, $\Psi _{2 ,1 }$ and $\Psi _{2 ,2 }$, for an open SL with  $n$=8. In this case the parity $P\bigl[\mathfrak{R}{\rm e}[\Psi _{\mu ,\nu }]\bigl]$ is given by  $(-1)^{\nu}$. See Eq. (\ref{SimmPolinUn}). Shown in grey at the background is the structure of the SL}
\end{center}
\end{figure}
In figures 3 and 4,  we plot the real parts of the resonant wave functions $\Psi _{1 ,1 }$,  $\Psi _{1 ,2 }$, $\Psi _{2 ,1 }$ and $\Psi _{2 ,2 }$, for  $n$=7 and $n$=8, respectively. Their symmetries correspond with those in (\ref{SimmPolinUn}). Notice that for $n$ odd, the wave function parity depends also on the subband index $\mu$.  How are the symmetries of the imaginary parts? To this purpose, let us explore the resonant wave function (\ref{OpenSLWF}), at points inside the SL. Let these points be $z_{1}=-L/2+l_c$ and $z_{n-1}=L/2-l_c$. At these points we have
\begin{eqnarray}
\Psi_{\mu,\nu}(z_1)=\left(\alpha + \beta^*-(\beta+\alpha^*)\frac{\beta_n^*}{\alpha_n^*}\right)a_o
\end{eqnarray}
and
\begin{eqnarray}
\Psi_{\mu,\nu}(z_1)=\left(\alpha_{n-1} \!+\! \beta_{n-1}^*-(\beta_{n-1}\!+\!\alpha_{n-1}^*)\frac{\beta_n^*}{\alpha_n^*}\right)a_o.\hspace{0.3in}
\end{eqnarray}
Taking into account that $\beta_n^*|_{E_{\mu,\nu}}=\beta U_{n-1}|_{E_{\mu,\nu}}=0$ and taking into account the identities\cite{Pereyra2002}
\begin{eqnarray}\label{nm1Identities1}
\alpha_{n-1}=\alpha_n\alpha^*-\beta_n\beta^*, \label{nm1Identities1}\\
\beta_{n-1}=-\alpha_n\beta+\beta_n\alpha,\label{nm1Identities2}
\end{eqnarray}
we obtain
\begin{eqnarray}
\Psi_{\mu,\nu}(z_1)=(\alpha + \beta^*)
a_o
\end{eqnarray}
and
\begin{eqnarray}
\Psi_{\mu,\nu}(z_{n-1})=(\alpha^* - \beta^*)\frac{1}{\alpha_n^*}
a_o,
\end{eqnarray}
whose real parts satisfy the relation
\begin{eqnarray}
\mathfrak{R}{\rm e}[\Psi_{\mu,\nu}(z_{n-1})]=\frac{1}{U_n}\frac{\alpha_R+\beta_R}{\alpha_R-\beta_R}\mathfrak{R}{\rm e}[\Psi_{\mu,\nu}(z_1)].
\end{eqnarray}
When $\beta_R=0$ and the norm symmetry requirement for symmetric SL wave functions, is met,\cite{symmetricnorm} we have
\begin{eqnarray}
\mathfrak{R}{\rm e}[\Psi_{\mu,\nu}(z_{n-1})]=\frac{1}{U_n}\mathfrak{R}{\rm e}[\Psi_{\mu,\nu}(z_1)].
\end{eqnarray}
This relation coincides with (\ref{OpenWFSymms1}) and agrees with the wave function symmetries, shown  in figures 3 and 4. On the other hand, the imaginary parts of $\Psi_{\mu,\nu}(z_1)$ and $\Psi_{\mu,\nu}(z_{n-1})$ satisfy the relation
\begin{eqnarray}\label{WFSimRelImP}
\mathfrak{I}{\rm m}[\Psi_{\mu,\nu}(z_{n-1})]=-\frac{1}{U_n}\mathfrak{I}{\rm m}[\Psi_{\mu,\nu}(z_1)]
\end{eqnarray}
\begin{figure}\label{WFopenSLn7ImP}
\begin{center}
\includegraphics [width=210pt]{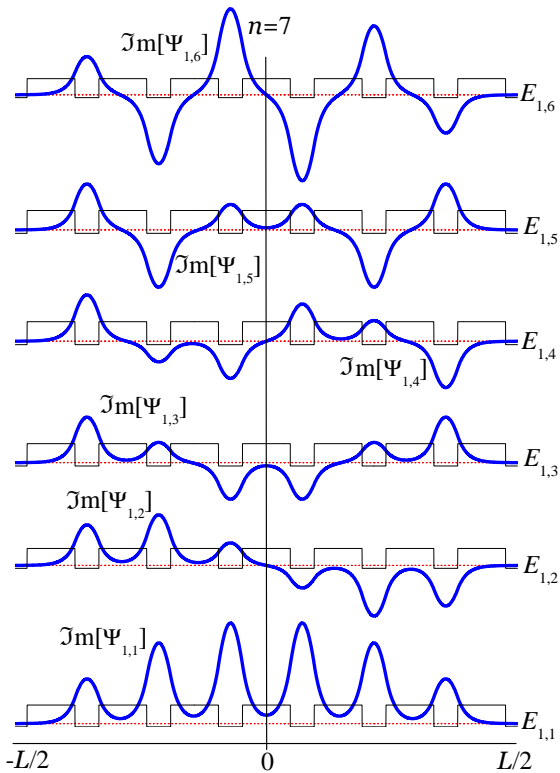}
\caption{Imaginary parts of the resonant wave functions $\Psi _{1 ,1 }$,  $\Psi _{1 ,2 }$, $\Psi _{1 ,3 }$ ... $\Psi _{1 ,6 }$, for an open SL with  $n$=7. Now  the parity $P\bigl[\mathfrak{I}{\rm m}[\Psi _{\mu ,\nu }]\bigl]$, according with  Eqs. (\ref{SimmPolinUn}), (\ref{WFSimRelImP}), and (\ref{WFopenSLn7ImAndReP}), is given by  $(-1)^{\nu+1}$. }
\end{center}
\end{figure}
In figure 5 we plot the imaginary parts of the resonant states in the first subband. They fulfill the symmetries of $-U_n$, as the last equation suggests. This relation together with the symmetry relation for the real part, imply that
\begin{eqnarray}
\Psi_{\mu,\nu}(L/2-l_c)=\frac{1}{U_n}\Psi_{\mu,\nu}^*(-L/2+l_c).
\end{eqnarray}
Therefore the resonant wave functions' space-inversion symmetries, of open and symmetric 1D superlattices, are given by
\begin{eqnarray}
\Psi_{\mu,\nu}(z)=\frac{1}{U_n}\Psi_{\mu,\nu}^*(-z).
\end{eqnarray}
or
\begin{eqnarray}\label{WFopenSLn7ImAndReP}
\Psi_{\mu,\nu}(z)\!=\!\Biggl\{\begin{array}{cc} (-1)^{\nu}\Psi_{\mu,\nu}^*(-z) & \text{for}\hspace{0.1in} n \hspace{0.1in}\text{even}\cr  & \cr (-1)^{\nu+\mu+1}\Psi_{\mu,\nu}^*(-z) & \text{for}\hspace{0.1in} n \hspace{0.1in}\text{odd}  \end{array}\Biggr.
\end{eqnarray}
This relation comprises all the results known, so far, for the resonant 1D SL wave-function symmetries. In the next sections we will derive symmetry relations  for superlattice eigenfunctions.

\section{Eigenfunctions' symmetries in bounded superlattices}

\begin{figure}\label{Fig6}
\begin{center}
\includegraphics [width=210pt]{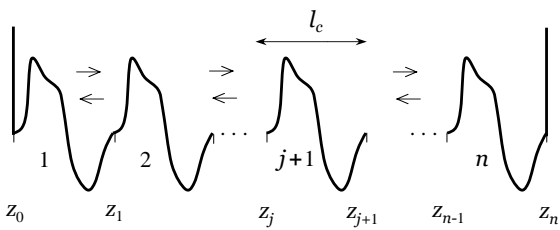}
\caption{Parameters of a bounded SL with exactly $n$ cells.}
\end{center}
\end{figure}
\begin{figure}\label{Fig7}
\begin{center}
\includegraphics [width=215pt]{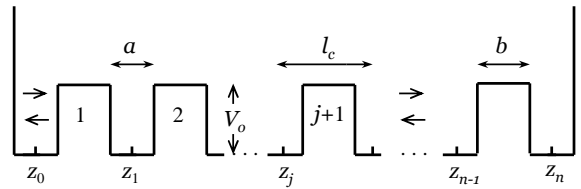}
\caption{Parameters of a bounded $n$-cells SL with extra half wells at the ends.}
\end{center}
\end{figure}

There is a family of  optoelectronic devices  where the active SLs are bounded systems. Generally, they are bounded by thick cladding layers, and the confining potentials are modeled with finite-height lateral barriers. In these systems, the SL eigenfunctions penetrate the lateral barriers and decay exponentially. The physics of this kind of systems, that we called quasi-bounded SLs,  is rather similar to that of the bounded SL by hard walls or completely confined SLs as were named in Ref. [\onlinecite{Pacher2007}]. In the completely confined SLs, the wave functions vanish at the walls and the eigenvalues and eigenfunctions equations are slightly simpler that for the quasi-bounded SLs by finite-height lateral walls. Let us consider now the  infinite walls case.

It was shown in Ref. [\onlinecite{Pereyra2005}] that the eigenvalues $E_{\mu,\nu}$ of a  bounded SL,  in the 1D one-mode approximation, are obtained from
\begin{eqnarray}\label{EIgEqBSLExactly1}
\alpha _{n}+\beta _{n}^{\ast }=\alpha _{n}^{\ast }+\beta_{n},
\end{eqnarray}
or equivalently, from
\begin{eqnarray}\label{EIgEqBSLExactly2}
U_{n-1}(\alpha_I-\beta_I)= 0,
\end{eqnarray}
when the distance between the hard walls is  $L=nl_c$, with exactly $n$ unit cells between the hard walls, as in figure 6. When the bounded SL is like in figure 7, with an extra half well at the ends, the  distance between the walls is $L=nl_c+a$, and the eigenvalues are obtained from
\begin{eqnarray}\label{EIgEqBSL}
\alpha _{n}e^{i k a}+\beta _{n}^{\ast }=\alpha _{n}^{\ast }e^{-i k a}+\beta_{n},
\end{eqnarray}
that can be written as
\begin{eqnarray}
U_n \sin ka + (\alpha_I \cos ka-\alpha_R \sin ka-\beta_I)U_{n-1}= 0,\hspace{0.2in}
\end{eqnarray}
where  $\alpha_I$ and $\beta_I$ are the imaginary parts of $\alpha$ and $\beta$, the elements (1,1) and (1,2) of the single-cell transfer matrix. We will see now an interesting change in the eigenfunction symmetries of these cases. Let us  discuss now the easiest case of SLs whose total length is $L=nl_c$.

\subsection{Eigenfunction symmetries of bounded SL with length $L=nl_c$}

\begin{figure}\label{Fig8}
\begin{center}
\includegraphics [width=210pt]{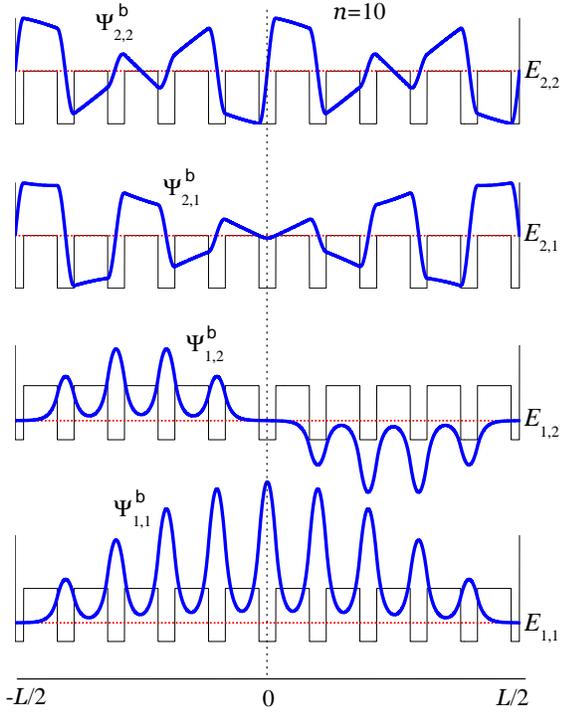}
\caption{Real parts of the resonant wave functions $\Psi _{1 ,1 }$,  $\Psi _{1 ,2 }$, $\Psi _{2 ,1 }$ and $\Psi _{2 ,2 }$, for a bounded SL with  length $L=nl_c$ and $n$=10. In this case the parity $P\bigl[\Psi _{\mu ,\nu }\bigl]$ is given by  $(-1)^{\nu+1}$. See Eq. (\ref{BWFSymExactly}).}
\end{center}
\end{figure}
In this case, according with the eigenvalue equation (\ref{EIgEqBSLExactly2}), we have, on one side, the eigenvalues determined by the zeros of the Chebyshev polynomial $U_{n-1}$. On the other, the vanishing of the factor $\alpha_I-\beta_I$ provides, as was shown in Ref. [\onlinecite{Pereyra2005}], the eigenvalues that correspond to the surface states. The contribution of surface states to optical transitions will depend on the relation $V_w/V_o$, between the chadding barriers height $V_w$ and the SLs barrier height $V_o$. We will restrict the symmetry analysis to eigenfunctions that correspond to the eigenvalues that nullify the Chebyshev polynomial  $U_{n-1}$. When the length of the bounded SL is $L=nl_c$, the wave function at any point $-L/2\leq z \leq L/2$ is given by
\begin{eqnarray}
\Psi ^{b}(z,E) &\!=\!&A\left[(\alpha _{p}+\gamma _{p})\left( \alpha _{j}-\beta _{j}%
e^{i \theta_n}\right)\right.
 \nonumber \\
&&+\left.(\beta _{p}+\delta _{p})\left( \beta _{j}^{\ast }-\alpha
_{j}^*e^{i \theta_n}\right) \right],
\end{eqnarray}
where $A$ is a normalization constant and
\begin{eqnarray}
e^{i \theta_n}=\frac{\alpha _{n}+\beta _{n}^{\ast }}{\alpha _{n}^{\ast }+\beta_{n}}.
\end{eqnarray}
The eigenfunctions are obtained from
\begin{eqnarray}
\Psi _{\mu ,\nu }^{b}(z)=\Psi ^{b}(z,E_{\mu ,\nu }).
\end{eqnarray}
\begin{figure}\label{Fig9}
\begin{center}
\includegraphics [width=210pt]{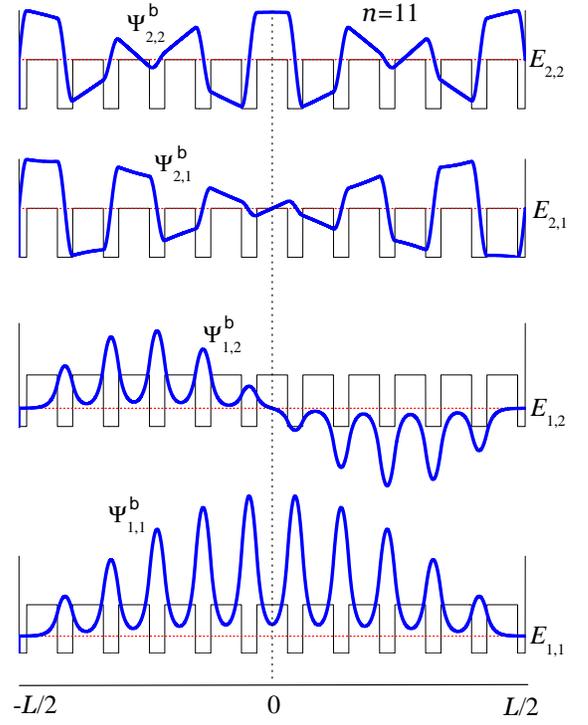}
\caption{Real parts of the resonant wave functions $\Psi _{1 ,1 }$,  $\Psi _{1 ,2 }$, $\Psi _{2 ,1 }$ and $\Psi _{2 ,2 }$, for a bounded SL with  length $L=nl_c$ and $n$=11. In this case the parity $P\bigl[\Psi _{\mu ,\nu }\bigl]$ is given by  $(-1)^{\mu+\nu}$. See Eq. (\ref{BWFSymExactly}).}
\end{center}
\end{figure}
It is easy to verify that the eigenfunctions satisfy the boundary conditions\cite{wallsBSL1}, i.e. that $\Psi _{\mu ,\nu }^{b}(-L/2)=\Psi _{\mu ,\nu}^{b}(L/2)$=0. To unveil the wave functions' symmetries, we can evaluate the eigenfunctions at any other two points, symmetric with respect to the origin. For example, at $z=z_1=-L/2+l_c$ and at $z=z_{n-1}=L/2-l_c$. At these points, $\alpha_p=\delta_p=1$ and $\beta_p=\gamma_p=0$. At $z_1$,  $\alpha_j=\alpha$ and $\beta_j=\beta$, while at  $z_{n-1}$,  $\alpha_j=\alpha_{n-1}$ and $\beta_j=\beta_{n-1}$. Notice that, because of the eigenvalues equation, $e^{i\theta_n}\bigr|_{E_{\mu,\nu}}\!=1$. Thus
\begin{eqnarray}
\Psi_{\mu,\nu}^b(z_1)=A (\alpha-\beta+\beta^*-\alpha^*).
\end{eqnarray}
Similarly, we have
\begin{eqnarray}
\Psi_{\mu,\nu}^b(z_{n-1})=A(\alpha_{n-1}-\beta_{n-1}+\beta^*_{n-1}-\alpha^*_{n-1}).
\end{eqnarray}
Using the identities (\ref{nm1Identities1}) and (\ref{nm1Identities2})
we obtain
\begin{eqnarray}
\Psi_{\mu,\nu}^b(z_{n-1})=A(\alpha_n+\beta_n^*)(\alpha^*+\beta-\beta^*-\alpha),
\end{eqnarray}
which means
\begin{eqnarray}
\Psi_{\mu,\nu}^b(L/2-l_c)=-(\alpha_n+\beta_n^*)\Psi_{\mu,\nu}^b(-L/2+l_c).
\end{eqnarray}
Since the imaginary part of $\alpha_n+\beta_n^*$, according with the eigenvalues equation is zero, we are left with
\begin{eqnarray}
(\alpha_n\!+\!\beta_n^*)|_{E_{\mu,\nu}}\!=\!\left[U_n\!-\!\alpha^*U_{n-1}\right]\Bigr|_{E_{\mu,\nu}}\!=U_n\Bigr|_{E_{\mu,\nu}}.\hspace{0.2in}
\end{eqnarray}
Therefore
\begin{eqnarray}
\Psi_{\mu,\nu}^b(L/2-l_c)=-U_n\Psi_{\mu,\nu}^b(-L/2+l_c),.
\end{eqnarray}
Taking into account the Chebyshev polynomial symmetries in equation (\ref{SimmPolinUn}), we end up with
\begin{eqnarray}\label{BWFSymExactly}
\Psi_{\mu,\nu}^b(z)\!=\!\Biggl\{\begin{array}{cc} (-1)^{\nu+1}\Psi_{\mu,\nu}^b(-z) & \text{for}\hspace{0.1in} n \hspace{0.1in}\text{even}\cr  & \cr (-1)^{\nu+\mu}\Psi_{\mu,\nu}^b(-z) & \text{for}\hspace{0.1in} n \hspace{0.1in}\text{odd}  \end{array}\Biggr.
\end{eqnarray}
The eigenfunctions $\Psi _{1 ,1 }^b$,  $\Psi _{1 ,2 }^b$, $\Psi _{2 ,1 }^b$ and $\Psi _{2 ,2 }^b$, for $n$=10 and $n$=11, respectively, are shown in figures 8 and 9. It is clear from these graphs that the parity symmetries, of equation (\ref{BWFSymExactly}), are fulfilled.

\subsection{Eigenfunction symmetries for bounded SLs with length $L=nl_c+a$}

It was shown, also in Ref. [\onlinecite{Pereyra2005}], that for this kind of system, where all wells have the same width, the wave function at any point $-L/2-a/2\leq z \leq L/2+a/2$ is given by
\begin{eqnarray}
\Psi ^{b}(z,E) &\!=\!&a_oe^{i k a/2}\left[(\alpha _{p}+\gamma _{p})\left( \alpha _{j}-\beta _{j}%
e^{i \theta_n}\right)\right.
 \nonumber \\
&&+\left.(\beta _{p}+\delta _{p})\left( \beta _{j}^{\ast }-\alpha
_{j}^*e^{i \theta_n}\right) \right],
\end{eqnarray}
with $a_o$ a normalization constant and
\begin{eqnarray}
e^{i \theta_n}=\frac{\alpha _{n}+\beta _{n}^{\ast }e^{-i k a}}{\alpha _{n}^{\ast }e^{-i k a}+\beta_{n}}.
\end{eqnarray}
\begin{figure}\label{Fig10}
\begin{center}
\includegraphics [width=210pt]{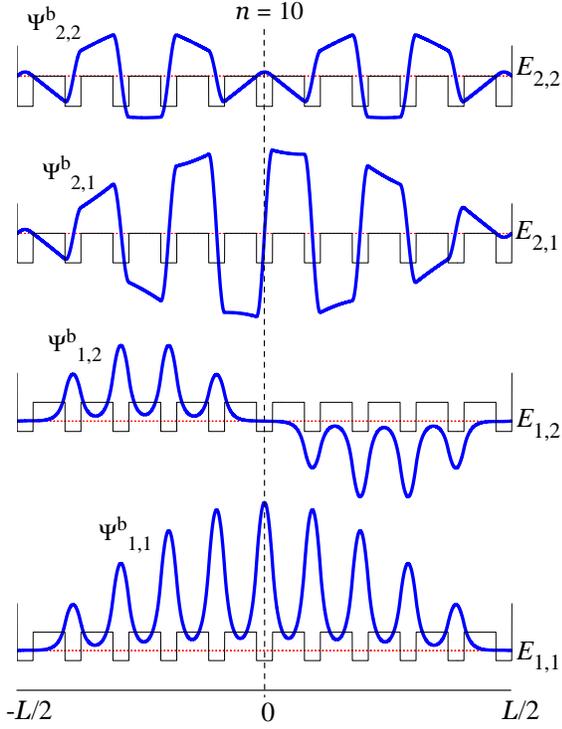}
\caption{Real parts of the resonant wave functions $\Psi _{1 ,1 }$,  $\Psi _{1 ,2 }$, $\Psi _{2 ,1 }$ and $\Psi _{2 ,2 }$, for a bounded SL with  length $L=nl_c+a$ and $n$=10. In this case the parity $P\bigl[\Psi _{\mu ,\nu }\bigl]$ is given by  $(-1)^{\nu+\mu}$. See Eq. (\ref{BWFSym}).}
\end{center}
\end{figure}
\begin{figure}\label{Fig11}
\begin{center}
\includegraphics [width=210pt]{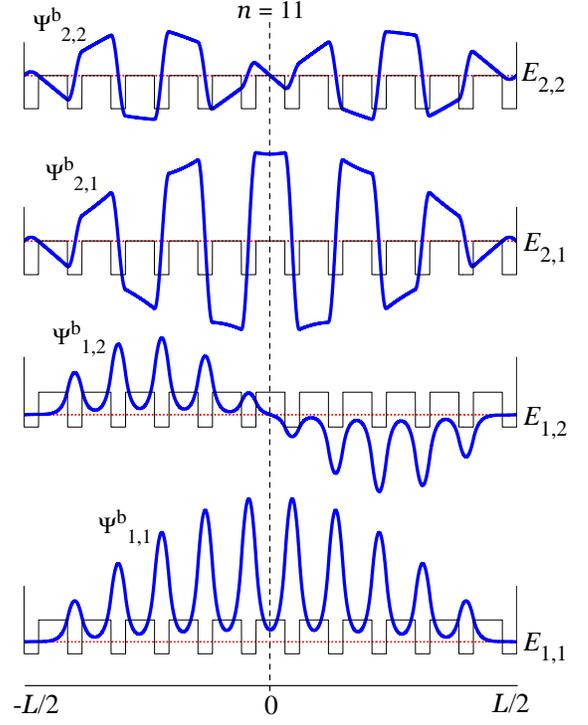}
\caption{Real parts of the resonant wave functions $\Psi _{1 ,1 }$,  $\Psi _{1 ,2 }$, $\Psi _{2 ,1 }$ and $\Psi _{2 ,2 }$, for a bounded SL with  length $L=nl_c+a$ and $n$=11. In this case the parity $P\bigl[\Psi _{\mu ,\nu }\bigl]$ is given by  $(-1)^{\nu+1}$. See Eq. (\ref{BWFSym}).}
\end{center}
\end{figure}
Also in this case, it is easy to verify\cite{wallsBSL2} that the eigenfunctions vanish at the walls, i.e. that $\Psi _{\mu ,\nu }^{b}(-L/2-a/2)=\Psi _{\mu ,\nu }^{b}(L/2+a/2)$=0. Again, to unveil the wave functions' symmetries, we can evaluate the eigenfunctions at any two symmetric points, say at $z_o=-L/2$ and at $z_n=L/2$. At these points, $\alpha_p=\delta_p=1$ and $\beta_p=\gamma_p=0$. At $z_o$,  $\alpha_j=1$ and $\beta_j=0$, while, at  $z_n$,  $\alpha_j=\alpha_n$ and $\beta_j=\beta_n$. Thus
\begin{eqnarray}
\Psi^b(z_o,E_{\mu,\nu})=a_oe^{i k a/2}\left (1-e^{i \theta_n}\right ),
\end{eqnarray}
and
\begin{eqnarray}
\Psi^b(z_o,E_{\mu,\nu})=a_oe^{i k a/2}\left (\alpha_n +\beta_n^*-(\beta_n+\alpha_n^*)e^{i \theta_n}\right ).\hspace{0.1in}
\end{eqnarray}
Using the eigenvalues equation (\ref{EIgEqBSL}), we have
\begin{eqnarray}
\Psi^b(z_o,E_{\mu,\nu})=a_o(e^{i k a/2}-e^{-i k a/2}).
\end{eqnarray}
and
\begin{eqnarray}
\Psi^b(z_n,E_{\mu,\nu})=a_o\frac{e^{-i k a/2}-e^{i k a/2}}{\alpha_n^*e^{-i k a}+ \beta_n}.
\end{eqnarray}
This means that
\begin{eqnarray}
\Psi_{\mu,\nu}^b(L/2)=-\frac{1}{\alpha_n^*e^{-i k a}+ \beta_n}\Psi_{\mu,\nu}^b(-L/2).
\end{eqnarray}
It turns out that the factor $(\alpha_n^*e^{-i k a}+ \beta_n)\bigl |_{E=E_{\mu,\nu}}$, whose imaginary part vanishes because of the eigenvalues equation, takes the  values
\begin{eqnarray}
\frac{1}{\alpha_n^*e^{-i k a}+\beta_n}\Bigl |_{E=E_{\mu,\nu}}\!=\!\Biggl\{\begin{array}{cc} (-1)^{\nu} & \text{for}\hspace{0.1in} n \hspace{0.1in}\text{odd}\cr & \cr  (-1)^{\nu+\mu+1} & \text{for}\hspace{0.1in} n \hspace{0.1in}\text{even}  \end{array}\Biggr.
\end{eqnarray}
Therefore, the eigenfunction symmetries of bounded SL with length $L=nl_c+a$, like in figure 7,  are the following
\begin{eqnarray}\label{BWFSym}
\Psi_{\mu,\nu}^b(z)\!=\!\Biggl\{\begin{array}{cc} (-1)^{\nu+1}\Psi_{\mu,\nu}^b(-z) & \text{for}\hspace{0.1in} n \hspace{0.1in}\text{odd}\cr  & \cr (-1)^{\nu+\mu}\Psi_{\mu,\nu}^b(-z) & \text{for}\hspace{0.1in} n \hspace{0.1in}\text{even}  \end{array}\Biggr.
\end{eqnarray}
In figures 10 and 11,  we plot the eigenfunctions $\Psi _{1 ,1 }^b$,  $\Psi _{1 ,2 }^b$, $\Psi _{2 ,1 }^b$ and $\Psi _{2 ,2 }^b$, for $n$=10 and $n$=11, respectively. It is clear from these graphs that the spacial inversion eigenfunction symmetries, described in equation (\ref{BWFSym}), are fulfilled.

It is worth noticing the important differences between  the eigenfunction symmetries of a SL with length $L=nl_c$ and those of a SL with length $L=nl_c+a$. Let us now discuss the more realist case of SL bounded by lateral barriers with finite height, as shown in Fig. 12.

\section{Eigenfunctions symmetries in quasi-bounded superlattices}

Confined superlattices are found in the active zone of SL lasers, where  periodic structures are grown within cladding layers or reflection layers. The blue emitting SLs studied by Nakamura et al. are some examples.  For a bounded SL like in figure 12, the eigenvalues equation was given in Ref. [\onlinecite{Pereyra2005}]  as
\begin{eqnarray}
h_wU_n+f_wU_{n-1}= 0,
\end{eqnarray}
with
\begin{equation}
h_{w}=\frac{q_{w}^{2}-k^{2}}{2q_{w}k}\sin ka+\cos ka.
\end{equation}
and
\begin{eqnarray}
f_{w} &=&\frac{q_{w}^{2}-k^{2}}{2q_{w}k}(\alpha _{I}\cos ka-\alpha
_{R}\sin ka) -\alpha _{R}\cos ka \nonumber \\
&&-\alpha _{I}\sin ka-\beta _{I}\frac{q_{w}^{2}+k^{2}%
}{2q_{w}k}
\end{eqnarray}
Here $q_w=\sqrt{2 m^*(V_w-E)/\hbar^2}$ is the wave number in the side barriers. It  is also easy to verify that  the wave function at any point $z$, in the $j+1$ cell, is given by
\begin{eqnarray}
\Psi^{qb}(z,E) &\!\!=\!\!&\!\frac{a_o}{2k}\Bigl[\Bigl((\alpha_{p}\!+\!\gamma
_{p})\alpha
_{j}\!+\!(\beta_{p}\!+\!\delta_{p})\beta_{j}^{\ast}\Bigr)e^{\!i k a/2}(k\!-\!iq_{w}
)\Bigr. \nonumber \\ &\!\!+\!\!&\Bigl.\Bigl((\alpha
_{p}\!+\!\gamma_{p})\beta_{j}\!+\!(\beta_{p}\!+\!\delta _{p})\alpha _{j}^{\ast
}\Bigr) e^{\!-i k a/2}(k\!+\!iq_{w})\Bigr],\nonumber \\
 \label{e.15}
\end{eqnarray}
with $a_o$ a normalization constant and the matrices $\alpha_p$, ..., $\beta_j$ as defined before. The eigenfunctions are obtained also from
\begin{eqnarray}
\Psi _{\mu ,\nu }^{qb}(z)=\Psi ^{qb}(z,E_{\mu ,\nu }).
\end{eqnarray}
\begin{figure}
\begin{center}
\includegraphics [width=220pt]{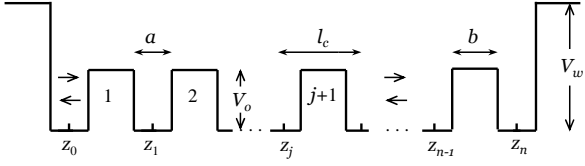}
\caption{Parameters of a quasi-bounded $n$-cells SL with lateral barrier height $V_w$=0.5783eV.}
\end{center}
\end{figure}
\begin{figure}
\begin{center}
\includegraphics [width=210pt]{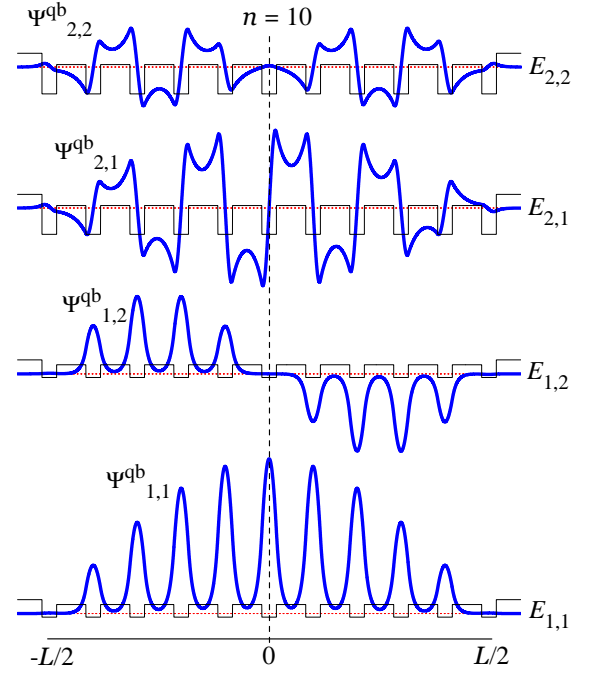}
\caption{Real parts of the resonant wave functions $\Psi _{1 ,1 }$,  $\Psi _{1 ,2 }$, $\Psi _{2 ,1 }$ and $\Psi _{2 ,2 }$, for a quasi-bounded SL with  length $L=nl_c+a$ and $n$=10. In this case the parity $P\bigl[\Psi _{\mu ,\nu }\bigl]$ is given by  $(-1)^{\nu+\mu}$. See Eq. (\ref{QBWFSym}).}
\end{center}
\end{figure}
At the borders of the lateral barriers, i.e. at $-L/2$ and at $L/2$, the eigenfunctions are non-zero. The relation between the wave functions at these points is
\begin{eqnarray}\label{QBWFSymEnds}
\Psi_{\mu,\nu}^{qb}(L/2)&\!=\!&\Bigl[\frac{k^2\!+\!q_w^2}{2kq_w}(\alpha_{nR}\sin ka+\alpha_{nI}\cos ka) \nonumber \\ &\!+\!&\frac{k^2\!-\!q_w^2}{2kq_w}\beta_{nI}  \Bigr]\Psi_{\mu,\nu}^{qb}(\!-\!L/2\!). \hspace{0.2in}
\end{eqnarray}
\begin{figure}
\begin{center}
\includegraphics [width=210pt]{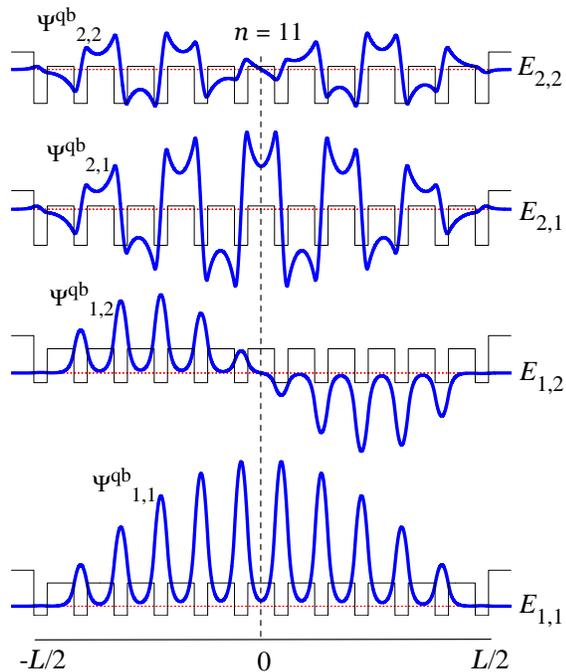}
\caption{Real parts of the resonant wave functions $\Psi _{1 ,1 }$,  $\Psi _{1 ,2 }$, $\Psi _{2 ,1 }$ and $\Psi _{2 ,2 }$, for a quasi-bounded SL with  length $L=nl_c+a$ and $n$=11. In this case the parity $P\bigl[\Psi _{\mu ,\nu }\bigl]$ is given by  $(-1)^{\nu+1}$. See Eq. (\ref{QBWFSym}).}
\end{center}
\end{figure}
The factor multiplying the function $\Psi_{\mu,\nu}^b(\!-\!L/2)$ turns out to be also +1 or -1, as in the previous cases, with  symmetries similar to those of $(\alpha_n^*e^{-i k a}+ \beta_n)\bigl |_{E=E_{\mu,\nu}}$, in the bounded SL with length $L=nl_c+a$. Similar relations can be obtained for any other pair of symmetric points. If we choose, for example, the points at $z=z_0=-L/2+a/2$ and $z=z_n=L/2-a/2$, we will find the relation
\begin{eqnarray}
\Psi_{\mu,\nu}^{qb}(L/2-a/2)&\!=\!&\Bigl[ (\beta_{nI}-\alpha_{nI})\frac{k\sin ka/2-q_w\cos ka/2}{k\cos ka/2+q_w\sin ka/2}\nonumber \\ && + \alpha_{nR} \Bigr]\Psi_{\mu,\nu}^{qb}(\!-\!L/2+a/2). \hspace{0.2in}
\end{eqnarray}
with the same symmetries as in (\ref{QBWFSymEnds}).
Thus, the quasi-bounded SL eigenfunctions symmetries, at any point $z$,  are given by
\begin{eqnarray}\label{QBWFSym}
\Psi_{\mu,\nu}^{qb}(z)\!=\!\Biggl\{\begin{array}{cc} (-1)^{\nu+1}\Psi_{\mu,\nu}^{qb}(-z) & \text{for}\hspace{0.1in} n \hspace{0.1in}\text{odd}\cr  & \cr (-1)^{\nu+\mu}\Psi_{\mu,\nu}^{qb}(-z) & \text{for}\hspace{0.1in} n \hspace{0.1in}\text{even}  \end{array}\Biggr.
\end{eqnarray}
In figures 13 and 14,  we plot the eigenfunctions $\Psi _{1 ,1 }^{qb}$,  $\Psi _{1 ,2 }^{qb}$, $\Psi _{2 ,1 }^{qb}$ and $\Psi _{2 ,2 }^{qb}$, for $n$=10 and $n$=11, respectively.  Again the spatial inversion symmetries described in (\ref{QBWFSym}) are realized in the specific examples. Notice that the symmetries in quasi-bounded superlattices are the same as in the completely bounded SLs.

\section{conclusions}

We presented here a comprehensive derivation of the eigenfunction symmetries for open, bounded and quasi-bounded 1D periodic structures. These properties are fundamental in the theory of periodic systems and for applications, particularly for optical transition calculations. We have shown that the eigenfunctions are either even or odd, and we have found that these eigenfunction parities are fully determined by the number of unit cells $n$ and by the subband and intrasubband indices $\mu$ and $\nu$.


\begin{references}
\bibitem{symmetricnorm} It is known that for symmetric square barrier $\beta_R=0$.
\bibitem{wallsBSL1} To evaluate the wave functions at the walls of the bounded SL with length $L=nl_c$ one has to use, for $z=z_0=-L/2$, $\alpha_p=\delta_p=1$, $\beta_p=\gamma_p=0$, $\alpha_j=1$ and $\beta_j=0$; Replacing these numbers, we have $\Psi_{\mu,\nu}=0$. For $z=z_n=L/2$, we use $\alpha_p=\delta_p=1$, $\beta_p=\gamma_p=0$, $\alpha_j=\alpha_n$ and $\beta_j=\beta_n$. Replacing we have  $\Psi_{\mu,\nu}=A \left(\alpha_n-\beta_n+\beta_n^*-\alpha_n^* \right)$ which agrees with the function in the eigenvalues equation (\ref{EIgEqBSL}) that vanish for $E=E_{\mu,\nu}$.
\bibitem{wallsBSL2} To evaluate the wave functions at the walls of the bounded SL one has to use, for $z=-L/2-a/2$, $\alpha_p=e^{-ika/2},\delta_p=e^{ika/2}$, $\beta_p=\gamma_p=0$, $\alpha_j=1$ and $\beta_j=0$; for $z=L/2+a/2$, $\alpha_p=e^{ika/2},\delta_p=e^{-ika/2}$, $\beta_p=\gamma_p=0$, $\alpha_j=\alpha_n$ and $\beta_j=\beta_n$. In addition one has to use the eigenvalues equation (\ref{EIgEqBSL}) and the identity $\alpha_n\alpha_n^*-\beta_n\beta_n^*=1$.
\end{references}
\end{document}